\documentclass{article}
\usepackage{cite}
\usepackage{graphicx}
\usepackage{dcolumn}

\begin{document}

\date{}
\title{On the misinterpretation of conditionally-solvable quantum-mechanical
problems}
\author{Francisco M. Fern\'{a}ndez\thanks{%
fernande@quimica.unlp.edu.ar} \\
INIFTA, DQT, Sucursal 4, C.C 16, \\
1900 La Plata, Argentina}
\maketitle

\begin{abstract}
We apply the Frobenius (power-series) method to some simple exactly-solvable
and conditionally-solvable quantum-mechanical models with supposed physical
interest. We show that the supposedly exact solutions to radial eigenvalue
equations derived in recent papers are not correct because they do not
satisfy some well-known theorems. We also discuss the origin of the mistake
by means of the approach indicated above.
\end{abstract}

\section{Introduction}

\label{sec:intro}

Conditionally-solvable quantum-mechanical problems have been of great
interest during the last decades (see, for example, Turbiner's remarkable
review\cite{T16} and the references therein). However, the solutions to the
eigenvalue equations stemming from such models have been misinterpreted in a
wide variety of physical applications\cite{AF20,AF21,F21}.

In a recent paper, Mustafa\cite{M22a} derived apparently exact solutions to
an eigenvalue equation that is known to be conditionally solvable\cite
{AF20,AF21,F21}. For this reason, we deem it necessary to discuss Mustafa's
results in some detail. In section~\ref{sec:Phys_models} we outline
Mustafa's models. In section~\ref{sec:HO} we solve one of them, which is
actually exactly solvable, by means of the Frobenius (power-series) method.
In section~\ref{sec:CLH} we apply the Frobenius method to a
conditionally-solvable radial eigenvalue equation that contains Mustafa's
ones as particular cases. Finally, in section~\ref{sec:conclusions} we
summarize the main results and draw conclusions.

\section{Physical models}

\label{sec:Phys_models}

In this section we outline the eigenvalue equations for three models
discussed by Mustafa\cite{M22a}. The physical meaning of the parameters is
not relevant for present discussion and the interested reader is referred to
that paper.

For the ``KG-oscillator in cosmic string spacetime within KKT'' Mustafa\cite
{M22a} derived the radial equation
\begin{equation}
U^{\prime \prime }(r)+\left[ \tilde{\lambda}+\frac{1/4-\tilde{\gamma}^{2}}{%
r^{2}}-\tilde{\omega}^{2}r^{2}\right] U(r)=0,  \label{eq:HO_Mus}
\end{equation}
and obtained the eigenvalues
\begin{equation}
\tilde{\lambda}^{M}=2\tilde{\omega}\left( 2n_{r}+|\tilde{\gamma}|+1\right) ,
\label{eq:HO_lambda^M}
\end{equation}
where $n_{r}=0,1,\ldots $ is the radial quantum number.

By means of the change of variables $\rho =\tilde{\omega}^{1/2}r$ we obtain
\begin{equation}
F^{\prime \prime }(\rho )+\left[ \frac{\tilde{\lambda}}{\tilde{\omega}}+%
\frac{1/4-\tilde{\gamma}^{2}}{\rho ^{2}}-\rho ^{2}\right] F(\rho )=0.
\label{eq:HO}
\end{equation}

For the ``pseudo-confined PDM KG-oscillator in cosmic string spacetime
within KKT'' Mustafa\cite{M22a} derived the radial equation
\begin{equation}
U^{\prime \prime }(r)+\left[ \mathcal{E}+\frac{1/4-\tilde{\beta}^{2}}{r^{2}}-%
\tilde{\omega}^{2}r^{2}-\eta \tilde{a}r-\frac{\tilde{b}}{r}\right] U(r)=0,
\label{eq:CLH1_Mus}
\end{equation}
and obtained the eigenvalues
\begin{equation}
\mathcal{E}^{M}=2\tilde{\omega}\left( 2n_{r}+|\tilde{\beta}|+1\right) -\frac{%
\tilde{a}^{2}\eta ^{2}}{4\tilde{\omega}^{2}}.  \label{eq:CLH1_E^M}
\end{equation}
Curiously, these eigenvalues do not depend on $\tilde{b}$ in spite of the
fact that the eigenvalue equation already depends on this model parameter.
Well-known general theorems are useful for testing results; here, we can
resort to the celebrated Hellmann-Feynman theorem (HFT)\cite{G32,F39} that
in the present case states that
\begin{equation}
\frac{\partial \mathcal{E}}{\partial \tilde{b}}=\left\langle \frac{1}{r}%
\right\rangle >0,\;\frac{\partial \mathcal{E}}{\partial \eta }=\tilde{a}%
\left\langle r\right\rangle .  \label{eq:CLH1_HFT}
\end{equation}
Clearly, Mustafa's result cannot be correct because
\begin{equation}
\frac{\partial \mathcal{E}^{M}}{\partial \tilde{b}}=0,\;\frac{\partial
\mathcal{E}^{M}}{\partial \eta }=-\frac{\tilde{a}^{2}\eta }{2\tilde{\omega}%
^{2}}.  \label{eq:CLH1_HFT_M}
\end{equation}
Another unmistakable indication that equation (\ref{eq:CLH1_E^M}) is not
correct is that $\mathcal{E}^{M}$ does not yield the eigenvalues of the
Coulomb problem when $\tilde{\omega}=0$ and $\tilde{a}=0$.

The same change of variables indicated above yields
\begin{equation}
F^{\prime \prime }(\rho )+\left[ \frac{\mathcal{E}}{\tilde{\omega}}+\frac{%
1/4-\tilde{\beta}^{2}}{\rho ^{2}}-\rho ^{2}-\frac{\eta \tilde{a}}{\tilde{%
\omega}^{3/2}}\rho -\frac{\tilde{b}}{\tilde{\omega}^{1/2}\rho }\right]
F(\rho )=0.  \label{eq:CLH1}
\end{equation}

For the ``confined PDM KG-oscillator-III in cosmic string spacetime within
KKT'' Mustafa\cite{M22a} derived the radial eigenvalue equation
\begin{equation}
U^{\prime \prime }(r)+\left[ \tilde{\lambda}_{1}+\frac{1/4-\tilde{\gamma}%
_{1}^{2}}{r^{2}}-\tilde{\omega}_{1}^{2}r^{2}-2mAr-\frac{2B}{r}\right] U(r)=0,
\label{eq:CLH2_Mus}
\end{equation}
and obtained the eigenvalues
\begin{equation}
\tilde{\lambda}_{1}^{M}=2\tilde{\omega}_{1}\left( 2n_{r}+|\tilde{\gamma}%
_{1}|+1\right) -\frac{m^{2}A^{2}}{\tilde{\omega}_{1}^{2}}.
\label{eq:CLH2_lambda_1^M}
\end{equation}
In this case, the HFT states that
\begin{equation}
\frac{\partial \tilde{\lambda}_{1}}{\partial B}=2\left\langle \frac{1}{r}%
\right\rangle >0,\;\frac{\partial \tilde{\lambda}_{1}}{\partial A}%
=2m\left\langle r\right\rangle ,  \label{eq:CLH2_HFT}
\end{equation}
which are not satisfied by Mustafa's $\tilde{\lambda}_{1}^{M}$. Besides, $%
\tilde{\lambda}_{1}^{M}$ does not yield the eigenvalues of the Coulomb
problem when $\tilde{\omega}_{1}=0$ and $A=0$.

The change of variables $\rho =\tilde{\omega}_{1}^{1/2}r$ yields
\begin{equation}
F^{\prime \prime }(\rho )+\left[ \frac{\tilde{\lambda}_{1}}{\tilde{\omega}%
_{1}}+\frac{1/4-\tilde{\gamma}_{1}^{2}}{\rho ^{2}}-\rho ^{2}-\frac{2mA}{%
\tilde{\omega}_{1}^{3/2}}\rho -\frac{2B}{\tilde{\omega}_{1}^{1/2}\rho }%
\right] F(\rho )=0.  \label{eq:CLH2}
\end{equation}

It is clear from the analysis above that Mustafa's analytical expressions
for the energy in his equations (33) and (38) cannot be correct.
Consequently, all the physical conclusions derived from such equations, and
shown in Mustafa's figures 3 and 4, are based on wrong analytical
expressions for the eigenvalues $\mathcal{E}^{M}$ and $\tilde{\lambda}%
_{1}^{M}$.

The three radial eigenvalue equations outlined above are particular cases of
\begin{equation}
F^{\prime \prime }(\rho )+\left[ W+\frac{1/4-\gamma ^{2}}{\rho ^{2}}-\rho
^{2}-\frac{a}{\rho }-b\rho \right] F(\rho )=0,  \label{eq:CLH}
\end{equation}
where $a$ and $b$ are arbitrary real parameters. We are interested in
square-integrable solutions $F(\rho )$ that vanish at origin. Such solutions
only take place for particular values of the eigenvalue $W$ that we may
label as $W_{\nu ,\gamma }(a,b)$, $\nu =0,1,\ldots $, in such a way that $%
W_{\nu ,\gamma }<W_{\nu +1,\gamma }$.

From the HFT we conclude that
\begin{equation}
\frac{\partial W}{\partial a}=\left\langle \frac{1}{\rho }\right\rangle >0,\;%
\frac{\partial W}{\partial b}=\left\langle \rho \right\rangle >0.
\label{eq:HFT}
\end{equation}

\section{The exactly-solvable case}

\label{sec:HO}

The eigenvalue equation (\ref{eq:CLH}) is exactly solvable \textit{only}
when $a=b=0$. In order to obtain exact solutions for this particularly
simple case we resort to the well known Frobenius (power-series) method. If
we insert the ansatz
\begin{equation}
F(\rho )=\rho ^{s}\exp \left( -\frac{\rho ^{2}}{2}\right) \sum_{j=0}^{\infty
}c_{j}\rho ^{2j},\;s=\left| \gamma \right| +\frac{1}{2},
\label{eq:HO_ansatz}
\end{equation}
into the eigenvalue equation we find that the expansion coefficients $c_{j}$
satisfy the two-term recurrence relation
\begin{equation}
c_{j+1}=\frac{4j+2s-W+1}{2\left( j+1\right) \left( 2j+2s+1\right) }%
c_{j},\;j=0,1,\ldots .  \label{eq:HO_TTRR}
\end{equation}
For arbitrary values of $W$ the solution (\ref{eq:HO_ansatz}) is not square
integrable\cite{F21b}; however if we choose
\begin{equation}
W=W_{\nu ,\gamma }=2\left( 2\nu +|\gamma |+1\right) ,\;\nu =0,1,\ldots ,
\label{eq:HO_W}
\end{equation}
the series reduces to a polynomial and the resulting eigenfunctions are
square integrable. Note that Mustafa's result (\ref{eq:HO_lambda^M}) is
correct because $W=$ $\tilde{\lambda}^{M}/\tilde{\omega}$ when $\nu =n_{r}$.
The two-term recurrence relation thus takes the following simpler form
\begin{equation}
c_{j+1,\nu ,\gamma }=\frac{2\left( j-\nu \right) }{\left( j+1\right) \left(
2j+2s+1\right) }c_{j,\nu ,\gamma },  \label{eq:HO_TTRR_2}
\end{equation}
and the eigenfunctions become
\begin{equation}
F_{\nu ,\gamma }(\rho )=\rho ^{s}\exp \left( -\frac{\rho ^{2}}{2}\right)
\sum_{j=0}^{\nu }c_{j,\nu ,\gamma }\rho ^{2j}.  \label{eq:HO_eigenf}
\end{equation}

\section{Conditionally-solvable models}

\label{sec:CLH}

When one of the model parameters $a$ or $b$ is nonzero the eigenvalue
equation (\ref{eq:CLH}) is not exactly solvable. It is an example of
conditionally-solvable quantum-mechanical problems\cite{T16}. Mustafa\cite
{M22a} obtained the results outlined above from a dubious analysis of the
biconfluent Heun function. In this section we resort to the Frobenius method
because it is not only simpler and clearer but leaves no room for doubts.

In the general case we resort to the ansatz
\begin{equation}
F(\rho )=\rho ^{s}\exp \left( -\frac{b}{2}\rho -\frac{\rho ^{2}}{2}\right)
\sum_{j=0}^{\infty }c_{j}\rho ^{j},\;s=\left| \gamma \right| +\frac{1}{2},
\label{eq:CLH_ansatz}
\end{equation}
that leads to the three-term recurrence relation
\begin{eqnarray}
c_{j+2} &=&A_{j}c_{j+1}+B_{j}c_{j}=0,\;j=-1,0,1,\ldots ,\;c_{-1}=0,\;c_{0}=1,
\nonumber \\
A_{j} &=&\frac{a+b\left( j+s+1\right) }{\left( j+2\right) \left(
j+2s+1\right) },\;B_{j}=\frac{4\left( 2j+2s-W+1\right) -b^{2}}{4\left(
j+2\right) \left( j+2s+1\right) }.  \label{eq:CLH_TTRR}
\end{eqnarray}
In order to have a polynomial of degree $n$ we require that $c_{n}\neq 0$, $%
c_{n+1}=0$ and $c_{n+2}=0$, $n=0,1,\ldots $, that clearly leads to $c_{j}=0$
for all $j>n$. It follows from this condition that $B_{n}=0$ that yields
\begin{equation}
W=W_{\gamma }^{(n)}=2n+2s+1-\frac{b^{2}}{4}=2\left( n+|\gamma |+1\right) -%
\frac{b^{2}}{4}.  \label{eq:CHL_W^n}
\end{equation}
When $W=W_{\gamma }^{(n)}$ $B_{j}$ takes the simpler form

\begin{equation}
B_{j}=\frac{2\left( j-n\right) }{\left( j+2\right) \left( j+2s+1\right) }.
\label{eq:CLH_B_j}
\end{equation}
Present expression for $W$ is not consistent with Mustafa's results (\ref
{eq:CLH1_E^M}) and (\ref{eq:CLH2_lambda_1^M}) unless $n=2n_{r}$. However,
this discrepancy is irrelevant because neither $\mathcal{E}^{M}$, $\tilde{%
\gamma}_{1}^{M}$ or $W^{(n)}_{\gamma}$ are the eigenvalues of the
corresponding radial equations\cite{AF20,AF21,F21}. Note that $W_{\gamma
}^{(n)}$ also fails to satisfy the HFT (\ref{eq:HFT}).

The most important point is that, in order to obtain such particular
polynomial solutions, we need a second condition $c_{n+1}(a,b)=0$ already
omitted by Mustafa\cite{M22a}. Since $c_{j}(a,b)$ is a polynomial function
of degree $j$ in each model parameter we conclude that we have $n+1$ roots $%
a^{(n,i)}(b)$, $i=1,2,\ldots ,n+1$, for a given value of $b$, or $%
b^{(n,i)}(a)$ for each value of $a$. It can be proved that all the roots are
real\cite{AF20,CDW00}. The exact polynomial solutions for a given value of $b
$ are
\begin{equation}
F_{\gamma }^{(n,i)}(\rho )=\rho ^{s}\exp \left( -\frac{b}{2}\rho -\frac{\rho
^{2}}{2}\right) \sum_{j=0}^{n}c_{j}^{(n,i)}\rho ^{j}.  \label{eq:CLH_eigenf}
\end{equation}
The meaning of this kind of solutions has been discussed extensively in
recent papers\cite{AF20,AF21,F21}. However, in order to make present one
sufficiently self contained and convince the reader that our results are
correct we show some examples in what follows.

When $n=0$ we obtain
\begin{equation}
c_{1}(a,b)=\frac{a+bs}{2s}=0,  \label{eq:c_1(a,b)=0}
\end{equation}
and the exact solution
\begin{equation}
F_{\gamma }^{(0)}(\rho )=\rho ^{s}\exp \left( -\frac{b}{2}\rho -\frac{\rho
^{2}}{2}\right) ,  \label{eq:F^(0)}
\end{equation}
of equation (\ref{eq:CLH}) with $W=W_{\gamma }^{(0)}$ and $a=a_{\gamma
}^{(0)}(b)=-bs$. Note that $F_{\gamma }^{(0)}(\rho )$ is the ground state of
a model given by $a=a_{\gamma }^{(0)}(b)$.

When $n=1$ the second condition reads
\begin{equation}
c_{2}(a,b)=\frac{a^{2}+ab\left( 2s+1\right) +b^{2}s\left( s+1\right) -4s}{%
4s\left( 2s+1\right) }=0,  \label{eq:c_2(a,b)=0}
\end{equation}
with roots

\begin{equation}
a_{\gamma }^{(1,1)}(b)=\frac{\sqrt{b^{2}+16s}-b\left( 2s+1\right) }{2}%
,\;a_{\gamma }^{(1,2)}(b)=-\frac{\sqrt{b^{2}+16s}+b\left( 2s+1\right) }{2},
\label{eq:a^(1,i)}
\end{equation}
from which we obtain
\begin{eqnarray}
F_{\gamma }^{(1,1)}(\rho ) &=&\rho ^{s}\exp \left( -\frac{b}{2}\rho -\frac{%
\rho ^{2}}{2}\right) \left( 1+\frac{\sqrt{b^{2}+16s}-b}{4s}\rho \right) ,
\nonumber \\
F_{\gamma }^{(1,2)}(\rho ) &=&\rho ^{s}\exp \left( -\frac{b}{2}\rho -\frac{%
\rho ^{2}}{2}\right) \left( 1-\frac{\sqrt{b^{2}+16s}+b}{4s}\rho \right) .
\label{eq:F^(1,i)}
\end{eqnarray}
Note that $F_{\gamma }^{(1,1)}(\rho )$ is the ground state of a model with $%
a=a_{\gamma }^{(1,1)}(b)$ while $F_{\gamma }^{(1,2)}(\rho )$ is the first
excited state for a model given by $a=a_{\gamma }^{(1,2)}(b)$, both for $%
W=W_{\gamma }^{(1)}$. Anybody can easily verify that the functions in
equations (\ref{eq:F^(0)}) and (\ref{eq:F^(1,i)}) are solutions of equation (%
\ref{eq:CLH}) under the conditions just indicated.

As already stated above, this kind of solutions has been extensively
discussed in recent papers\cite{AF20,AF21,F21}. We just want to point out
that the occurrence of the exact polynomial solutions (\ref{eq:CLH_eigenf})
requires a second condition (overlooked by Mustafa\cite{M22a}) that
restricts the values of the model parameters and that such solutions are not
the only ones\cite{AF20,AF21,F21}. Therefore, any physical conclusion
derived from eigenvalues like (\ref{eq:CLH1_E^M}) and (\ref
{eq:CLH2_lambda_1^M}) are meaningless. In particular, Mustafa's expressions
for the energy in his equations (33) and (38) cannot be correct.
Consequently, all the physical conclusions derived from such equations, and
shown in Mustafa's figures 3 and 4, are based on wrong analytical
expressions for the eigenvalues $\mathcal{E}^{M}$ and $\tilde{\lambda}%
_{1}^{M}$.

\section{Conclusions}

\label{sec:conclusions}

We have already shown that the exact results obtained by Mustafa\cite{M22a}
for the eigenvalue equations (\ref{eq:CLH1_Mus}) and (\ref{eq:CLH2_Mus}) are
not correct because they are not exactly solvable. Those equations are
conditionally solvable and one obtains some particular polynomial solutions
for particular values of the model parameters determined by a second
condition overlooked by Mustafa. This conclusion also applies to another
paper by the same author where he apparently derived exact results for a
similar conditionally-solvable problem\cite{M22b}. Most physical conclusions
commonly derived from the polynomial solutions to conditionally-solvable
eigenvalue equations are meaningless\cite{AF20,AF21,F21}.


\begin{thebibliography}{99}
\bibitem{T16}  A. V. Turbiner, One-dimensional quasi-exactly solvable
Schrodinger equations, Phys. Rep. 642 (2016) 1-71. arXiv:1603.02992
[quant-ph].

\bibitem{AF20}  P. Amore and F. M. Fern\'{a}ndez, On some conditionally
solvable quantum-mechanical problems, Phys. Scr. 95 (2020) 105201.
arXiv:2007.03448 [quant-ph].

\bibitem{AF21}  P. Amore and F. M. Fern\'{a}ndez, An ubiquitous three-term
recurrence relation, J. Math. Phys. 62 (2021) 032106. arXiv:2110.14526
[quant-ph].

\bibitem{F21}  F. M. Fern\'{a}ndez, A most misunderstood
conditionally-solvable quantum-mechanical model, Ann. Phys. 434 (2021)
168645. arXiv:2109.11545 [quant-ph].

\bibitem{M22a}  O. Mustafa, PDM Klein-Gordon oscillators in cosmic string
spacetime in magnetic and Aharonov-Bohm flux fields within the Kaluza-Klein
theory, Ann. Phys. 440 (2022) 168857.

\bibitem{G32}  P. G\"{u}ttinger, Das Verhalten von Atomen im magnetischen
Drehfeld, Z. Phys. 73 (1932) 169-184.

\bibitem{F39}  R. P. Feynman, Forces in Molecules, Phys. Rev. 56 (1939)
340-343.

\bibitem{F21b}  F. M. Fern\'{a}ndez, On the singular harmonic oscillator,
2021. arXiv:2112.03693 [quant-ph].

\bibitem{CDW00}  M. S. Child, S-H. Dong, and X-G. Wang, Quantum states of a
sextic potential: hidden symmetry and quantum monodromy, J. Phys. A 33
(2000) 5653-5661.

\bibitem{M22b}  O. Mustafa, Confined Klein-Gordon o scillator from a
(2+1)-dimensional G\"{u}rses to a G\"{u}rses or a pseudo-G\"{u}rses
space-time backgrounds: Invariance and isospectrality, Eur. Phys. J. C 82
(2022) 82.
\end{thebibliography}
\end{document}